\documentclass[fleqn,usenatbib]{mnras}
\usepackage{natbib}
\usepackage{hyperref}
\usepackage{breakurl}
\usepackage{mathtools}
\usepackage{graphicx}
\usepackage{newtxtext,newtxmath}
\usepackage[T1]{fontenc}
\DeclareRobustCommand{\VAN}[3]{#2}
\let\VANthebibliography\thebibliography
\def\thebibliography{\DeclareRobustCommand{\VAN}[3]{##3}\VANthebibliography}

\usepackage{booktabs}
\usepackage{amsmath}
\usepackage{multirow}
\usepackage{moresize}

\usepackage{graphicx}	
\usepackage{amsmath}	

\newcommand \source{Swift~J1727.8-1613}

\title[X-ray Polarization, Timing, and Spectroscopy]{Bridging X-ray Polarization with Timing \& Spectroscopic Parameters of a galactic black hole: \source}

\author[Arka Chatterjee et al.]{
Arka Chatterjee,$^{1}$\thanks{E-mail: arka.chatterjee@ddn.upes.ac.in}
Sujoy Kumar Nath,$^{2}$\thanks{E-mail: sknath@mx.nthu.edu.tw}
Kaushik Chatterjee,$^{3}$
Broja G. Dutta,$^{4}$
Samar Safi-Harb,$^{5}$
\newauthor
Indranil Chattopadhyay,$^{6}$
Sudip K. Garain$^{7, 8}$
and Hsiang-Kuang Chang$^{2, 10}$
\\
$^{1}$Department of Physics, School of Advanced Engineering, UPES, Dehradun, 248007, India\\
$^{2}$Institute of Astronomy, National Tsing Hua University, Hsinchu 300044, Taiwan\\
$^{3}$South-Western Institute for Astronomy Research (SWIFAR), Yunnan University, Kunming, Yunnan 650500, People’s Republic of China\\
$^{4}$Department of Physics, Rishi Bankim Chandra College, Naihati, West Bengal 743165, India\\
$^{5}$Department of Physics \& Astronomy, Faculty of Science, University of Manitoba, Winnipeg, Manitoba, R3T 2N2, Canada\\
$^{6}$Aryabhatta Research Institute of Observational Sciences (ARIES), Manora Peak, Nainital, 263001, India\\
$^{7}$Department of Physical Sciences, IISER, Mohanpur, WB, 741246, India\\
$^{8}$Center of Excellence in Space Sciences, IISER, Mohanpur, WB, 741246, India\\
$^{9}$Department of Physics, National Tsing Hua University, Hsinchu 300044, Taiwan\\
}

\date{Accepted XXX. Received YYY; in original form ZZZ}

\pubyear{\the\year{}}

\begin{document}
\label{firstpage}
\pagerange{\pageref{firstpage}--\pageref{lastpage}}
\maketitle

\begin{abstract}
We report a correlated energy-dependent time lag and degree of polarization 
for \source~ during its 2023 outburst. The energy-dependent time lag is measured around the type-C quasi-periodic oscillations (QPO) observed by {\it IXPE} on 2023-09-07, while the degree of polarization is obtained from energy-resolved polarimetric measurements.
The Spearman correlation coefficient was found to be 0.8, with a null hypothesis 
probability of 4.2\%. Furthermore, the correlation value drops as the quality factor, or Q 
value, of the observed QPO frequencies decreases. The spectral properties of \source~ are analyzed 
using simultaneous {\it Insight}/HXMT data. Thereafter, we present model-independent 
theoretical arguments to show that processes other than inverse 
Comptonization also contribute to both the observed polarization and time lags. 
This correlation may therefore point to additional mechanisms contributing to the connection between the spectral, temporal, and polarimetric properties of black hole binaries in their hard state.
\end{abstract}




\section{Introduction}
Accretion onto black holes is the source of the most energetic photons observed in the universe
and provides an excellent testbed to explore the extreme gravitational and 
thermodynamic environment around black holes. Outbursting black hole X-ray binaries exhibit
spectral and temporal evolution during the outburst \citep{MR2006}. The variation in the timing
properties could be explored through the quasi-periodic oscillations (QPOs), time lags, and 
variability analysis at numerous energy bands \citep{BH1990, B1997, R1999, VN1997}. Spectral 
evolution in the X-ray domain is mostly attributed to the variation in size, electron temperature, and optical depth of the Compton cloud or corona \citep{ST1980, Ny1994, CT1995, Esin1997, Chatterjee2017a}, leading to the 
variation of the Comptonization process \citep{HT1995}, which determines the evolving X-ray spectral 
index ($\alpha$). The photon generation associated with the accreting black hole could be of thermal
\citep{SS73} or non-thermal \citep{LZ87, WGZ00} origin.

The connection between the spectral and temporal evolutions of the black hole binaries has been explored \citep{Poutanen01, B05, CNMM02, TF04, ST09, RN14, Nath2024} in the post {\it RXTE} \citep{BRS93} era. A consensus has been achieved to associate the long-term temporal variabilities with the spectral index ($\Gamma$) variation. {\it IXPE} \citep{W22} observed several galactic \citep{Marra2024, Steiner2024, Veledina2024, Ninoyu2025, Majumder2026} as well as supermassive black holes \citep{Pal2023, Kouch2024, Mondal2024, Gianolli2024, Pal2025, Kouch2025, Maksym2025} where coronal geometry and radiative components are investigated considering Kerr geometry around black holes. Further, {\it IXPE} opened new frontiers to explore the spectral-temporal-polarimetric connections during the accretion onto black holes.

 In the context of the spectral-temporal-polarimetric connections during the black hole accretion, 
 \source~emerges as an excellent candidate. \source~went into an outburst phase on 24th August 2023
 \citep{Negoro2023a, Negoro2023b}. \cite{Veledina2023} discovered the X-ray polarization of \source. Later, \cite{Ingram2024} showed the variable nature of the X-ray polarization of \source. Moreover, 
 the radio polarization of \source~ was observed to be aligned with the X-ray polarization, which could 
 indicate the origin of X-ray emission from the jet region. QPOs were observed in the hard state \cite{PP2023}, and later QPO evolution during the initial days of the outburst was  
 presented in Figure 2 of \cite{JMMA2025}, where the centroid frequency varied from 0.1 to 10 Hz during the first 50 days of the outburst. The variation of the low-frequency QPO centroid frequency spanned 
two orders of magnitude, suggesting a contracting characteristic length-scale of similar magnitude.
 Detailed spectral and timing studies using Insight-HXMT showed the evolution of type-C QPOs along with changes in the accretion geometry, suggesting a shrinking Comptonizing region during the state transition \cite{Chatterjee2024}.
 An evolving QPO time lag that transitions from hard to soft lag at around a QPO frequency of $\sim 1.2$\, Hz has been studied by multiple authors \cite{Bollemeijer2025, Nath2026}. \cite{Ninoyu2025} explored the short-term variability and low-frequency QPOs and \cite{Ingram2024} reported the broadband time-lags using {\it IXPE} data. The broadband lags that are presented in Fig. 6 of 
 \cite{Ingram2024} differ significantly from the energy-dependent time lags around the QPO 
 centroid frequency and remain unexplored in the previous literature. Thus, the time lags across the QPOs need to be examined further to understand the complex radiative mechanisms close to the event 
 horizon.

 This article is structured in the following way. First, we employed the {\it IXPE} data to obtain
 the degree, angle of polarization, and energy-dependent polarizations. We extracted timing products,
 such as power density spectra (PDSs), energy-dependent time lags, fractional r.m.s, and dynamic PDSs 
 from the {\it IXPE} datasets. Later, we explored the spectral properties of the \source~ using 
 {\it Insight}/HXMT. Finally, we discuss the results connecting the timing, spectral, and polarimetric 
 data in our discussion section. 

\begin{figure}
\vskip 0.5cm
  \centering
  \includegraphics[angle=0,width=0.5\textwidth,keepaspectratio=true]{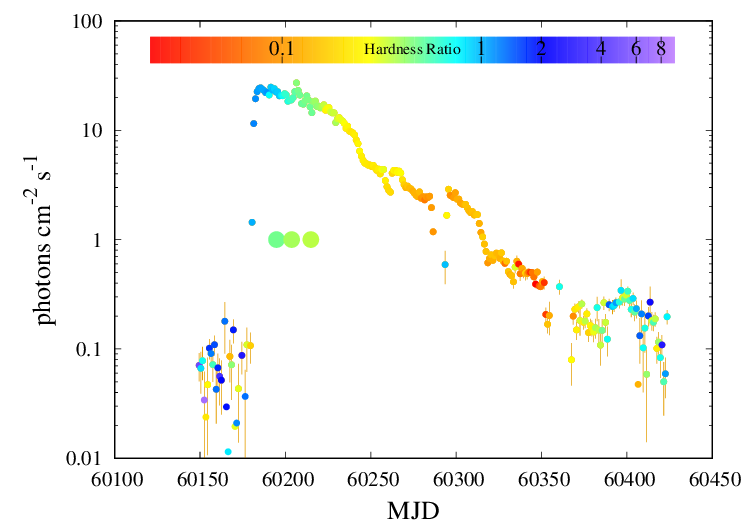}
	\caption{MAXI/GSC count rate is plotted with the MJD. The colorbar represents the hardness ratio ($4-20$ keV/$2-4$ keV) of GSC. $\bullet$ symbols represent the {\it IXPE} observations with distinct QPO features 
    over time.}
    \label{fig:hr}
\end{figure}

\section{Observations and Data Reduction} 
\label{sec:obs}
The variation of photon count rate, in the energy band of $2-20$ keV, obtained from {\it MAXI}/GSC, is presented
in Fig. \ref{fig:hr}. The colorbar represents the hardness ratio (HR) of the source. Considering the HR, 
color-coded {\it IXPE} observations are presented using the large filled circle symbol. Fig. \ref{fig:hr} shows that the brightness 
of the source was not maximum during the softest state. Rather, the softest possible state, achieved between MJD 60300 and 60350, corresponds to a luminosity almost two orders of magnitude lower than the maximum luminosity.  

\subsection{IXPE}
\label{sec:ixpe}
{\it IXPE} is a dedicated imaging satellite mission to measure X-ray polarization in the $2-8$ keV energy band \citep{Weisskopf2022}.
It consists of three identical polarization-sensitive gas-pixel detector units (DUs: \citet{Soffitta2021, Baldini2021}), each positioned at the focal plane of three mirror module assemblies \citep{Ramsey2022}.
We obtained cleaned and calibrated Level 2 data from the {\it IXPE} archive\footnote{\url{https://heasarc.gsfc.nasa.gov/docs/ixpe/archive}}
which were reduced with the \texttt{IXPEOBSSIM (v31.0.3)} software \citep{Baldini2022} and calibrated using \texttt{FTOOLS} tasks with the calibration database (CALDB v20241028). 
We selected a circular region of $96^{\prime\prime}$ radius centered on the source to extract the source events using the \texttt{xpselect} task.
Due to the source being very bright, we ignored any background contribution \citep{DiMarco2022} and used a `gray' filter in front of the detectors to reduce dead time effects by limiting the incident flux \citep{Ferrazzoli2020, Soffitta2021, Veledina2023}. Details of {\it IXPE} observations are presented in Table \ref{tab:ixpe_obs}.

\subsection{Insight-HXMT}
\label{sec:HXMT}

Insight-HXMT is equipped with three main payloads: the High Energy (HE: $20-250$ keV; \citealt{Liu2020}), Medium Energy (ME: $5-30$ keV; \citealt{Cao2020}), and Low Energy (LE: $1-15$ keV; \citealt{Chen2020}) 
detectors. The LE instrument consists of three Swept Charge Device (SCD) detectors, providing a total effective area of approximately $384$\,cm$^{2}$.
The ME instrument consists of $1728$ Silicon-PIN detectors, arranged into three detector groups, providing a total effective area of approximately $952$\,cm$^{2}$.
The HE instrument consists of $18$ NaI/CSI phoswich scintillation detectors, each providing an effective area of 
$286$\,cm$^{2}$. For our analysis, we retrieved level-1 data from the HXMT archive and generated level-2 cleaned data products 
suitable for analysis.
Data reduction was carried out using \texttt{HXMTDAS\footnote{\url{http://hxmt.org/index.php/usersp/dataan}} (v2.05)}, starting with the \texttt{hpipeline} task, 
which automatically executes a series of processing steps for all three instruments under predefined screening conditions. 
These criteria include: elevation angle $>$10$^{\circ}$, geomagnetic cutoff rigidity $>$8 GeV, pointing offset angle $<$0.04$^{\circ}$, 
and exclusion of time intervals within 600 s of South Atlantic Anomaly (SAA) passage. The cleaned event files were then used 
to extract spectra with \texttt{hespecgen}, \texttt{mespecgen}, and \texttt{lespecgen} tasks. 
The corresponding response files were generated using 
\texttt{herspgen}, \texttt{merspgen}, and \texttt{lerspgen} tasks. 
Instrumental background was modeled and subtracted using 
\texttt{hebkgmap}, \texttt{mebkgmap}, and \texttt{lebkgmap} for HE, ME, and LE, respectively. Details of {\it Insight}/HXMT observations are presented in Table \ref{tab:ixpe_obs}.
For spectral fitting, we grouped the data with \texttt{grppha} to ensure a minimum of 30 counts per bin for $\chi^2$ fit statistics in XSPEC.

\section{Data Analysis and Results}

\begin{table*}
\centering
\caption{List of \textit{IXPE} and \textit{Insight}/HXMT Observations Used} 
\label{tab:ixpe_obs}
\renewcommand{\arraystretch}{0.95}
\begin{tabular}{ccccc}
\toprule
Satellite & ObsID & Time & Obs Date & Exposure$^*$ \\
&&(MJD)&(yyyy-mm-dd)&(ks)\\
\midrule
{\it IXPE}& 02250901 & 60194.81 & 2023-09-07 & 19.0  \\
{\it Insight}/HXMT & P061433800807 & 60194.9 & 2023-09-07 & 11.4\\
\hline
{\it IXPE} & 02251001 & 60203.70 & 2023-09-16 & 36.9  \\
{\it Insight}/HXMT & P061433801704 & 60203.6 & 2023-09-16 & 11.4\\
\hline
{\it IXPE} & 02251101 & 60214.92 & 2023-09-27 & 21.1  \\
{\it Insight}/HXMT & P061433802808 & 60215.0 & 2023-09-27 & 10.3\\
\bottomrule
\end{tabular}
\begin{minipage}{10cm}
\vspace{2pt}
\footnotesize $^*$ Average effective exposure times of the DUs for IXPE observations, and total exposure times for \textit{Insight}/HXMT observations.
\end{minipage}
\end{table*}

\begin{figure*}
\vskip 0.5cm
  \centering
  \includegraphics[angle=0,width=\textwidth,keepaspectratio=true]{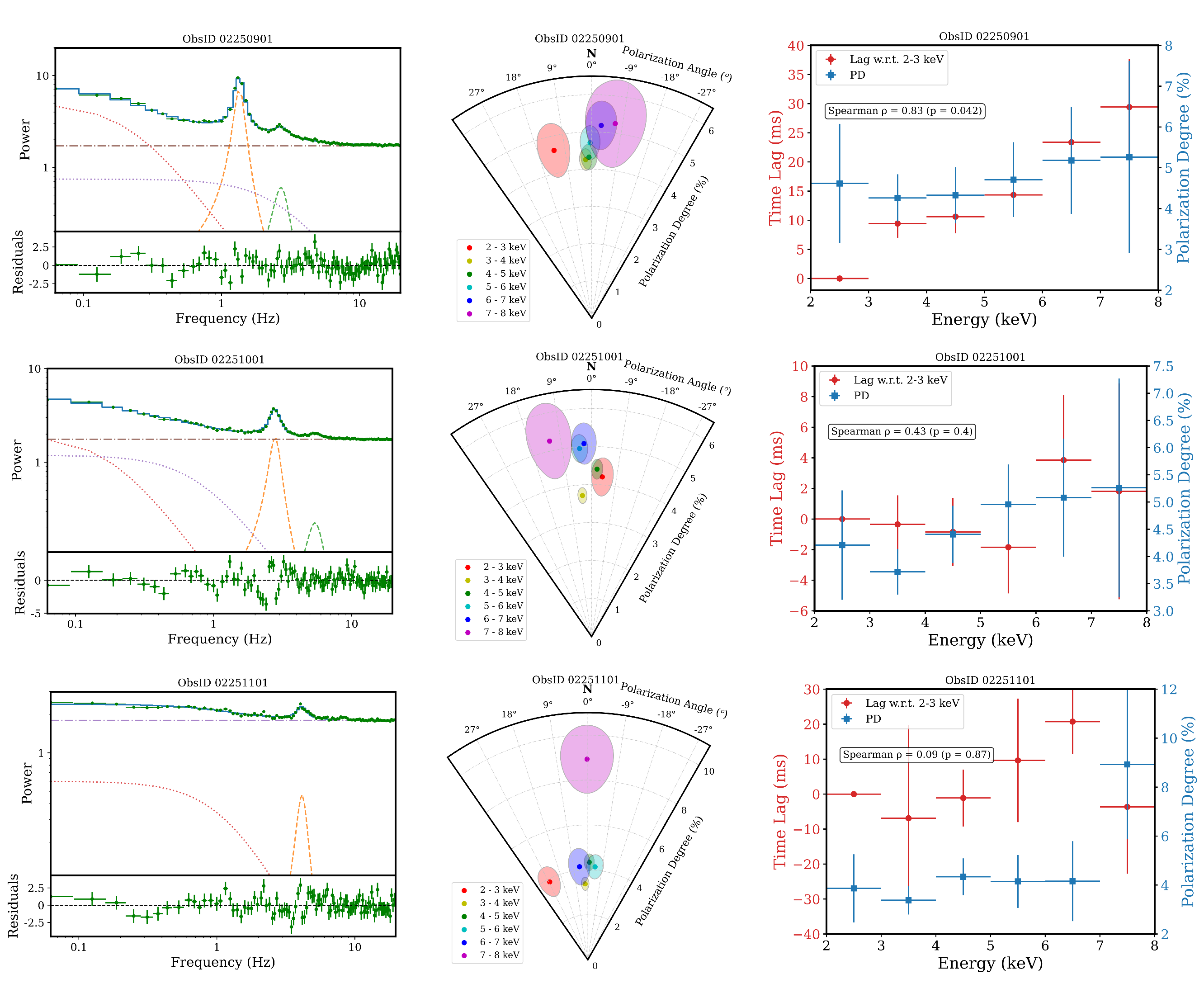}
	\caption{Left panel: Model-fitted Power Density Spectra (PDS) in the $2-8$ keV energy band for the three {\it IXPE} observations. The narrow Lorentzians corresponding to the QPO fundamental and harmonic peaks are shown in orange and green dotted lines, respectively. 
    Middle panel: Confidence contours between PA and PD as obtained from the PCUBE algorithm in the $2-3$, $3-4$, $4-5$, $5-6$, $6-7$, and $7-8$ keV energy bands. The shaded contours represent $1\sigma$ confidence intervals.
    Right panel: Correlation between time lag and PD in the $2-3$, $3-4$, $4-5$, $5-6$, $6-7$, and $7-8$ keV energy bands.}
    \label{fig:pds_papd_lag}
\end{figure*}

\begin{figure*}[t]
\vskip 0.5cm
  \centering
  \includegraphics[angle=0,width=\textwidth,keepaspectratio=true]{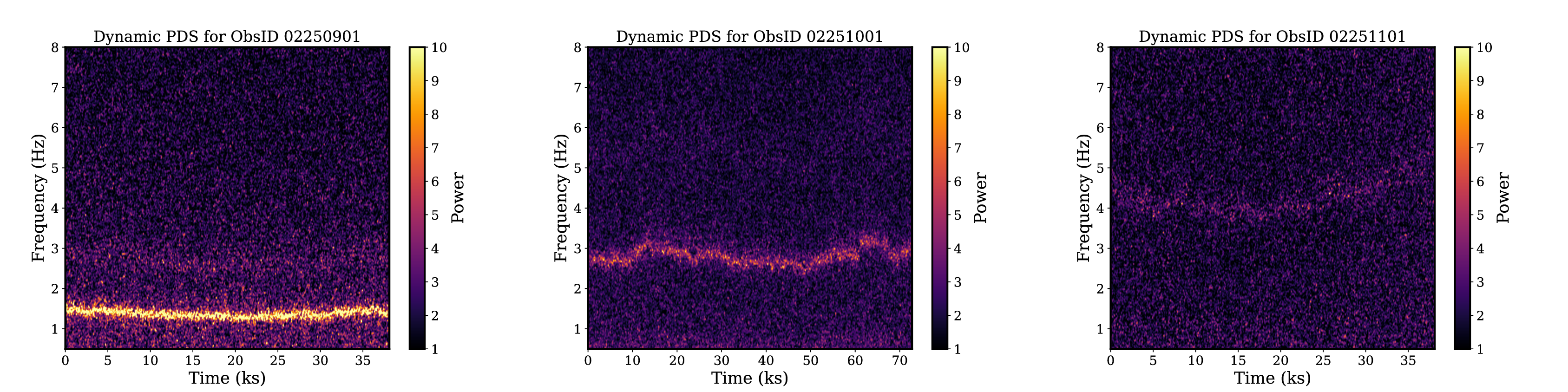}
	\caption{Dynamic PDS of the three {\it IXPE} observations are presented.}
    \label{fig:DPDS}
\end{figure*}

\begin{table*}
\centering
\caption{Timing and Polarization analysis Results. Q is the quality factor 
of the QPOs, PD represents the degree of polarization, and PA represents 
the angle of polarization.}
\label{tab:qpo_lag_pol}
\renewcommand{\arraystretch}{0.95}
\begin{tabular}{ccccccccc}
\toprule
ObsID & MJD & $\nu_{qpo}$ & Q & RMS & Energy Band & Time Lag & PD & PA \\
&&(Hz)&&(\%)&(keV)&(ms)&(\%)&(deg)\\
\midrule
\multirow{6}{*}{2250901} & \multirow{6}{*}{60194.8} & \multirow{6}{*}{$1.35 \pm 0.004$} & \multirow{6}{*}{$6.16 \pm 0.366$} & \multirow{6}{*}{$7.02 \pm 0.250$}
& 2--3 & $0.00 $ & $4.6 \pm 1.5$ & $12.6 \pm 9.2$ \\
& & & & & 3--4 & $9.4 \pm 2.4$ & $4.3 \pm 0.6$ & $2.2 \pm 3.9$ \\
& & & & & 4--5 & $10.6 \pm 2.8$ & $4.3 \pm 0.7$ & $0.9 \pm 5.0$ \\
& & & & & 5--6 & $14.3 \pm 3.6$ & $4.7 \pm 0.9$ & $0.5 \pm 5.6$ \\
& & & & & 6--7 & $23.4 \pm 5.6$ & $5.2 \pm 1.3$ & $-2.8 \pm 7.9$ \\
& & & & & 7--8 & $29.4 \pm 8.3$ & $5.3 \pm 2.4$ & $-6.9 \pm 15.2$ \\
\midrule
\multirow{6}{*}{2251001} & \multirow{6}{*}{60203.7} & \multirow{6}{*}{$2.79 \pm 0.008$} & \multirow{6}{*}{$5.12 \pm 0.226$} & \multirow{6}{*}{$8.74 \pm 0.235$}
& 2--3 & $0.00 $ & $4.2 \pm 1.0$ & $-3.9 \pm 6.9$ \\
& & & & & 3--4 & $-0.4 \pm 1.9$ & $3.7 \pm 0.4$ & $3.7 \pm 3.2$ \\
& & & & & 4--5 & $-0.8 \pm 2.2$ & $4.4 \pm 0.5$ & $-1.9 \pm 3.4$ \\
& & & & & 5--6 & $-1.8 \pm 3.0$ & $5.0 \pm 0.7$ & $3.7 \pm 4.4$ \\
& & & & & 6--7 & $3.8 \pm 4.2$ & $5.1 \pm 1.1$ & $2.3 \pm 6.4$ \\
& & & & & 7--8 & $1.8 \pm 7.1$ & $5.3 \pm 2.0$ & $12.2 \pm 11.1$ \\
\midrule
\multirow{6}{*}{2251101} & \multirow{6}{*}{60214.9} & \multirow{6}{*}{$4.14 \pm 0.044$} & \multirow{6}{*}{$5.11 \pm 0.797$} & \multirow{6}{*}{$6.89 \pm 0.647$}
& 2--3 & $0.00$ & $3.9 \pm 1.4$ & $26.3 \pm 11.9$ \\
& & & & & 3--4 & $-6.9 \pm 26.6$ & $3.4 \pm 0.6$ & $2.0 \pm 5.0$ \\
& & & & & 4--5 & $-1.1 \pm 8.1$ & $4.3 \pm 0.8$ & $-0.8 \pm 5.1$ \\
& & & & & 5--6 & $9.6 \pm 17.7$ & $4.2 \pm 1.1$ & $-4.5 \pm 8.9$ \\
& & & & & 6--7 & $20.7 \pm 9.2$ & $4.2 \pm 1.6$ & $5.2 \pm 11.6$ \\
& & & & & 7--8 & $-3.6 \pm 19.1$ & $8.9 \pm 3.0$ & $0.2 \pm 13.2$ \\
\bottomrule
\end{tabular}
\end{table*}

\begin{table}
\centering
\caption{Joint IXPE-HXMT Spectral fitting results}
\label{tab:spectral_results_pexrav}
\renewcommand{\arraystretch}{1.1}
\begin{tabular}{lccc}
\toprule
Epoch & 1 & 2 & 3 \\
\midrule
MJD & 60194.9 & 60203.6 & 60215.0 \\

\midrule
$N_{\rm H}$ ($\times10^{22}$ cm$^{-2}$) 
& $0.88^{+0.10}_{-0.08}$ 
& $0.63^{+0.05}_{-0.04}$
& $0.69^{+0.06}_{-0.05}$ \\

\midrule
\multicolumn{4}{c}{\textbf{DiskBB}} \\
\midrule
$T_{\rm in}$ (keV) 
& $0.46^{+0.01}_{-0.01}$ 
& $0.55^{+0.02}_{-0.01}$ 
& $0.66^{+0.02}_{-0.02}$ \\

Norm$_{\rm diskbb} (\times 10^3)$ 
& $133^{+26}_{-19}$ 
& $31^{+6}_{-4}$ 
& $12^{+2}_{-2}$ \\

\midrule
\multicolumn{4}{c}{\textbf{Power-law}} \\
\midrule
$\Gamma_{\rm PL}$ 
& $2.06^{+0.05}_{-0.04}$ 
& $2.27^{+0.05}_{-0.04}$ 
& $2.39^{+0.06}_{-0.04}$ \\

Norm$_{\rm PL}$ 
& $9.7^{+2.5}_{-1.7}$ 
& $13.6^{+3.6}_{-2.3}$ 
& $16.9^{+4.7}_{-3.5}$ \\

\midrule
\multicolumn{4}{c}{\textbf{Pexrav}} \\
\midrule

$\Gamma_{\rm pexrav}$ 
& $1.64^{+0.03}_{-0.02}$ 
& $2.19^{+0.03}_{-0.02}$ 
& $2.50^{+0.04}_{-0.03}$ \\

$E_{\rm cut}$ (keV) 
& $23.2^{+0.9}_{-0.7}$ 
& $33.9^{+2.1}_{-1.5}$ 
& $74.2^{+24.5}_{-11.9}$ \\

$rel_{refl}$ 
& $0.08^{+0.05}_{-0.03}$ 
& $0.33^{+0.04}_{-0.03}$ 
& $0.51^{+0.11}_{-0.08}$ \\

$Fe_{abund}$ 
& $1.23^{+2.21}_{-0.76}$ 
& $1.19^{+0.38}_{-0.24}$ 
& $0.91^{+0.28}_{-0.17}$ \\

Norm$_{\rm pexrav}$ 
& $28.5^{+1.97}_{-1.46}$ 
& $49.0^{+2.87}_{-2.61}$ 
& $59.2^{+6.35}_{-3.83}$ \\

\midrule
$\chi^{2}$ 
& 2238.41 
& 2422.43 
& 1986.85 \\

d.o.f. 
& 1875 
& 1875 
& 1875 \\

$\chi^{2}_{\rm red}$ 
& 1.19 
& 1.29 
& 1.06 \\

\bottomrule
\end{tabular}
\vspace{2mm}

\end{table}

\subsection{Energy-Dependent Polarization}
To estimate the energy dependent polarization properties, we used the model-independent \texttt{PCUBE} algorithm \citep{Kislat2015} of the \texttt{XPBIN} task to construct polarization cubes in the $2-3$, $3-4$, $4-5$, $5-6$, $6-7$, and $7-8$ keV energy bands and determined the normalized Stokes parameters (Q/I \& U/I) and the polarization degree (PD) and polarization angle (PA) from them.
The variation of the PD and PA in different energy bands for the three {\it IXPE} observations has been shown in Table \ref{tab:qpo_lag_pol}.
From the table, we can see that PD increases roughly with energy in all three observations. 
In the first two observations, the PD was $\sim4.6\%$ and $\sim4.2\%$, respectively, in the $2-3$ keV energy band, and increased to $\sim5.3\%$ (for both observations) in the $7-8$ keV energy band.
However, on the third observation, the PD was $\sim3.9\%$ in the $2-3$ keV energy band and increased to $\sim8.9\%$ in the $7-8$ keV energy band.
The PA decreased monotonically during the first observation, decreasing from $\sim13^\circ$ in the $2-3$ keV band to $\sim-7^\circ$ in the $7-8$ keV energy band.
The PA shows a roughly decreasing trend during the third observation, also decreasing from $\sim26^\circ$ in the $2-3$ keV band to $\sim-4^\circ$ in the $5-6$ keV band, while increasing to $\sim5^\circ$ in the $6-7$ keV band and decreasing to $\sim0^\circ$ in the $7-8$ keV band.
However, during the second observation, PA increases from $\sim-4^\circ$ in the $2-3$ keV band to $\sim12^\circ$ in the $7-8$ keV band. Figure \ref{fig:pds_papd_lag} shows the energy-dependent PA variation in the second column, and the variation of the degree of polarization with respect to the energy is presented 
in the third column.

\subsection{Temporal Analysis}

\subsubsection{Power Density Spectra (PDS)}

We constructed the power density spectrum (PDS) of the {\it IXPE} observations by combining all the DUs in 
$2-3$, $3-4$, $4-5$, $5-6$, $6-7$, and $7-8$ keV energy bands with the \texttt{Stingray (v2.1)} software \citep{Huppenkothen2019,bachetti2024_2,bachetti2024}.
We selected the time resolution to be 0.025\,s and divided the total duration of the observations into intervals of 16\,s.
We generated a PDS for each interval, and the individual PDSs were averaged to generate a single PDS, which was geometrically rebinned.
The PDS was normalised following the prescription of \citet{Leahy1983}.
Multiple narrow peaks were seen in the PDSs of all of the observations, indicative of a QPO.
To model the PDSs, a constant model was used to account for the Poisson noise, zero-centered Lorentzian profiles were used to account for the broadband noise, and narrow Lorentzian profiles were used to account for the fundamental QPO and its harmonics \citep{Nowak2000, Belloni2002}.
The power density spectra of the three observations and the best-fit models are shown in the leftmost column of Fig. \ref{fig:pds_papd_lag}.
The parameters for the fundamental QPO are mentioned in Table \ref{tab:qpo_lag_pol}.
A strong QPO was detected at $\sim 1.35$\,Hz in the first observation, with a $Q$-value of $\sim6.1$ and an RMS variability of $\sim7\%$, showing that this is a Type-C QPO \citep{Casella2005}.
The QPO frequency increased to $\sim2.79$\,Hz in the second observation and to $\sim4.14$\,Hz in the third observation, as the source evolved from hard to soft states.
The fundamental QPO frequency does not show any energy dependence, as also observed by \citep{Ninoyu2025}.

\subsubsection{Dynamic PDS}

To understand the rapid variation of the QPO frequency, we performed a dynamic power density spectrum analysis of each observation.
For this, we used lightcurves with a time resolution of $0.025$\,s corresponding to a Nyquist frequency of $20$\,Hz.
The lightcurves were divided into segments of $32$\,s duration, which sets the minimum resolvable frequency at $\sim0.03$\,Hz, which is sufficient to track QPO evolution.
For each segment, the PDS was computed with Leahy normalization and stacked sequentially to construct the dynamic PDS.
Figure \ref{fig:DPDS} shows the dynamic PDS for the three observations.
As seen from the leftmost panel of Fig. \ref{fig:DPDS}, the QPO centroid frequency remains almost stable at $\sim1.6$\,Hz throughout the first observation.
However, as seen in the middle panel, the QPO centroid frequency 
increases from the first observation, 
becomes unstable, and shows random transitions between $\sim2.8-3.2$\,Hz during the second observation.
The QPO frequency further increases in the third observation, and shows a clear transition between $\sim4-5$\,Hz as seen in the rightmost panel.
From the figure, it can also be observed that the QPO becomes weaker and broader with time, suggesting a transition toward a softer or less variable accretion state.
The dynamic PDS thus shows that the underlying oscillatory process responsible for the QPO becomes progressively less coherent and undergoes stochastic changes, indicating rapid changes in the inner accretion flow.
It also demonstrates the capability of \textit{IXPE} to probe fast accretion-driven variability in black hole X-ray binaries.

\subsubsection{Energy-Dependent Time Lag}
To measure the energy-dependent time lag, we constructed the cross-spectra in the  $3-4$, $4-5$, $5-6$, $6-7$, and $7-8$ keV energy bands with respect to the $2-3$ keV energy band \citep{vanderklis1987}. These cross-spectra were computed with a time bin size of $0.04$\,s, and then averaged over $10$\,s segments.
The frequency-dependent time-lag spectra are then calculated by dividing the argument of the complex cross-spectra by $2\pi\nu$ \citep{vaughan1997}.
From the time-lag spectra, we calculated the time-lag at the QPO frequency ($\nu_0$) by averaging the time-lag values over the frequency range $\nu_0 \pm \frac{\Delta\nu}{2}$ \citep{reig2000}, where $\Delta\nu$ is the FWHM of the QPO peak.
As per general convention, a positive value of the time-lag signifies hard lag, i.e., hard band photons lagging behind soft band photons, and a negative value signifies soft lag, i.e., soft band photons lagging behind hard band photons.
The obtained energy-dependent time lag values for the three observations are listed in Table \ref{tab:qpo_lag_pol}.
From the table, we find that during the first observation, the time lag was positive for all energy bands for the QPO frequency ($\nu_{\rm QPO}$) = 1.35 Hz, and it increases monotonically from $\sim9.4$\,ms in the $3-4$ keV energy band to $\sim29.4$\,ms in the $7-8$ keV energy band.
However, during the latter two observations, no such monotonic evolution of time lag with energy can be seen, and we find soft lag for the $\nu_{\rm QPO}$ 2.79 Hz and 4.14 Hz.

\subsection{Time Lag vs Degree of Polarization Relationship}
The variation of the time lag and the PD with energy is shown in the rightmost column of Figure \ref{fig:pds_papd_lag}.
From the figure, we can see that although the time lags seem to be correlated with the PD in the first observation, they do not seem to be correlated in the later observations.
To assess the degree of correlation among them, we calculated the Spearman correlation index between them.
The estimated values are shown inside the respective figures.
As we can see, the Spearman index ($\rho$) is $0.83$ for the first observation with a small p-value of 4.2\%, which indicates a strong positive correlation between time lag and PD during this observation. A p-value less than 5\% is widely accepted to negate the null hypothesis probability. 
However, the correlations deteriorate during the later observations, with Spearman $\rho = 0.43$ in the second observation, indicating a weak correlation, to Spearman $\rho = 0.09$ in the last observation, indicating the time lag and PD are uncorrelated during the last observation.   

\begin{figure*}
\vskip 0.5cm
  \centering
  \includegraphics[angle=0,width=\textwidth,keepaspectratio=true]{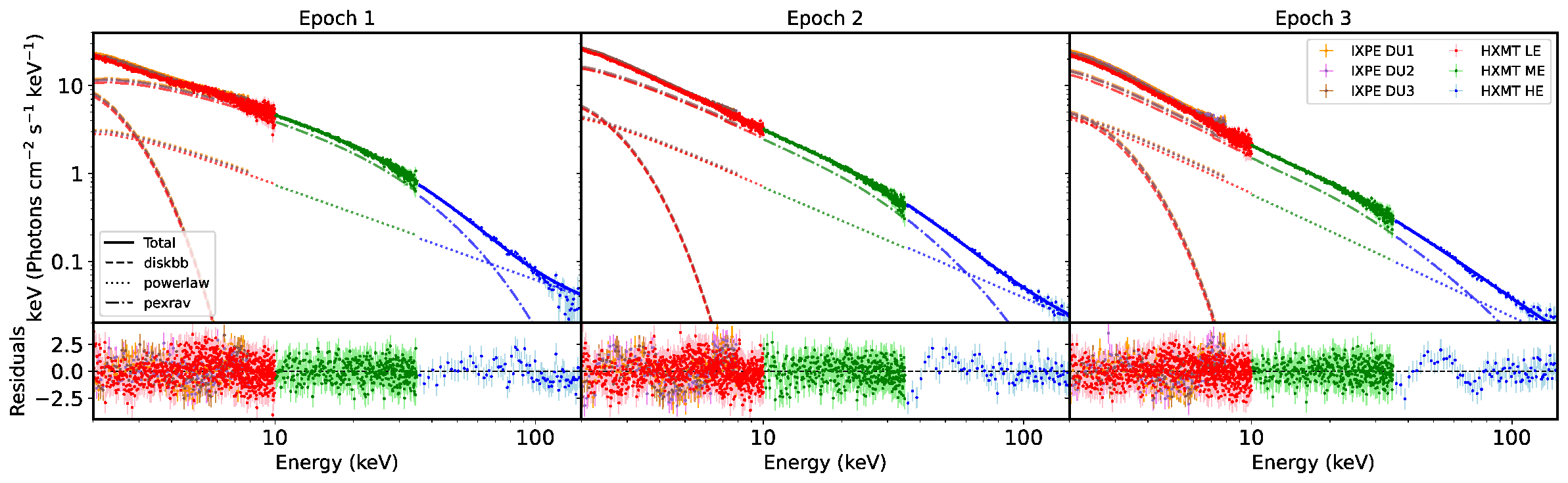}
	\caption{Simultaneous spectral fit of \textit{IXPE} Stokes \textit{I} spectra in $2-8$ keV (DU 1 in orange, DU 2 in purple, DU 3 in brown) and the three \textit{Insight}/HXMT instruments (LE: $2-10$ keV in red, ME: $10-35$ keV in green, and HE: $35-150$ keV in blue) with \texttt{tbabs(diskbb+powerlaw+pexrav)} model for the three epochs.}
    \label{}
\end{figure*}

\subsection{Spectral Analysis}

To determine the broadband spectral nature of the source during the {\it IXPE} epochs, we used data from the Insight-HXMT satellite contemporaneous with the {\it IXPE} observations.
We used data from all three instruments of the HXMT simultaneously (LE: $2-10$ keV, ME: $10-35$ keV, and HE: $35-110$ keV) to fit broadband spectra in the $2-110$ keV energy band.
Initially, we model the spectra as the sum of a disk-blackbody and a power-law, modified by interstellar absorption.
The photoelectric absorption cross sections were set to \texttt{vern} \citep{verner1996} and the relative abundances of elements in the ISM were set to \texttt{wilm} \citep{wilm2000}.
A multiplicative constant is used to account for the cross-calibration differences between LE, ME, and HE.
While the overall fit captures the broadband continuum reasonably well, systematic residuals are evident across all three instruments. 
In particular, a distinct high-energy rollover above $\sim70$ keV is observed in the HE spectrum, which the simple power-law component fails to reproduce.
This suggests that a more physically motivated Comptonization model is required to properly describe the hard X-ray emission.
We replaced the phenomenological power-law with the more physically motivated thermal Comptonization model \texttt{nthcomp} \citep{Z96,zycki1999} that describes the continuum spectra from the Compton up-scattering of soft-seed photons in a hot electron corona.
Next, we modeled the \textit{IXPE} Stokes \textit{I} spectra of all the DUs in the $2-8$ keV energy band simultaneously with the three \textit{Insight}/HXMT instruments.
We used an absorbed \texttt{disk-blackbody} plus \texttt{power-law} along with the \texttt{pexrav} model to take care of the reflection bump in the $20-30$ keV band. Variation of spectral parameters over the three epochs
are presented in Table \ref{tab:spectral_results_pexrav}.

\begin{figure*}
\vskip 0.5cm
  \centering
  \includegraphics[angle=0,width=\textwidth,keepaspectratio=true]{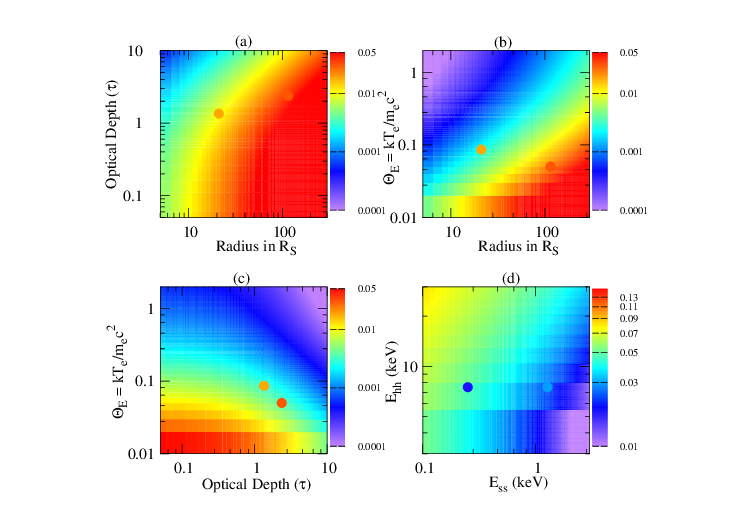}
	\caption{Based on equation \ref{eq:lag}, the parameter spaces of the thermodynamic variables contributing to the time lag are explored. The colorbar represents the positive lag magnitude. Panel (a) shows the time-lag map for optical depth $\tau$ and coronal radius ($R$) variation. Other parameters, such as $E_{hh}=100.0$ keV, $E_{ss}=1.0$ keV, and $\Theta= 0.05$ are 
    considered. Panel (b) represents the lag parameter space of coronal radius ($R$) and electron
    temperature ($\Theta_E$) keeping $\tau =0.05$, $E_{hh}=100.0$ keV, and $E_{ss}=1.0$ keV. Panel 
    (c) shows the variation of time lag with respect to $\tau =0.05$ and $R=40~R_g$ considering $\tau =0.05$, 
    $E_{hh}=100.0$ keV, and $E_{ss}=1.0$ keV. Lastly, panel (d) presents the time lag for the energy 
    enhancement considering $\tau =0.05$, $R=40~R_g$, and $\Theta_E = 0.5$. $\bullet$ symbols 
    represent the spectral parameters from the first two observations.}
    \label{fig:lag-param}
\end{figure*}

\section{Discussions}
\label{sec:discussion}

\subsection{Degree of Polarization and Time Lag}
\cite{Fano1957, M1961, Dolan1967, BS1987, NP1993, PV1993, Poutanen1994, Matt2004, Mcnamara2009, SK2010, BKM2017} extensively explored 
polarization (P) in the presence of Compton scattering, accretion disk geometry, and curved 
space-time. In general, the Comptonization-induced degree of polarization (P) is expected to 
be a function of energy (E) and time (t). The functional form of P could be written as
\begin{equation}
P(E,t) \propto f\left[\theta, \tau, \mathrm{t}_\mathrm{c}(\mathrm{E}_\mathrm{h}, \mathrm{E}_\mathrm{s})\right], 
\end{equation}
where $\theta$ is the angle between the incident and scattered photon, $\tau$ is the 
optical depth of the electron cloud, $\mathrm{t}_\mathrm{c}(\mathrm{E}_\mathrm{h}, \mathrm{E}_\mathrm{s})$ 
is the time lag due to Compton scattering, where $\mathrm{E}_\mathrm{h}$ \& $\mathrm{E}_\mathrm{s}$ are the energies of the hard and soft photons, respectively. Again, the time lag for a homogeneous 
Compton cloud of radius R and optical depth $\tau$ could be written as follows \citep{Payne1980, Nandi2021}

\begin{equation}
\mathrm{t}_\mathrm{c}(\mathrm{E}_\mathrm{h}, \mathrm{E}_\mathrm{s}) = \frac{R}{c(1+\tau)}\frac{ln(\mathrm{E}_\mathrm{h}/\mathrm{E}_\mathrm{s})}{ln\left[1+4\Theta(1+4\Theta)\right]},
\label{eq:lag}
\end{equation}
where $\Theta = \frac{kT_e}{m_ec^2}$, is the dimensionless electron temperature, 
c is the speed of light in vacuum. Eq. \ref{eq:lag} can lead to negative 
lags for $\mathrm{E}_\mathrm{h}/\mathrm{E}_\mathrm{s}<1$. This is, of course, 
an example of mathematical possibility denied by physical impossibility, as the 
Compton scattering would always require some amount of time. Using the 
spectral parameters, such as $\tau$, 
$\Theta$, $\mathrm{E}_\mathrm{h}$, $\mathrm{E}_\mathrm{s}$, and $R$, one can  
construct the parameter spaces for various time lags. The parameter spaces, 
using Eq. \ref{eq:lag}, are presented in Fig. \ref{fig:lag-param}. The 
color-bar in Fig. \ref{fig:lag-param} corresponds to the positive time lags 
observed between the lowest and the highest energy bands of the first two 
{\it IXPE} observations.  
Let us consider that the average number of scatterings suffered 
by a photon is $N_{scat}$ to reach the higher energy of 
$\mathrm{E}_\mathrm{h}$ from a lower energy level of $\mathrm{E}_\mathrm{s}$. 
So, 
\begin{equation}
(\mathrm{E}_\mathrm{h}/\mathrm{E}_\mathrm{s}) = \mathrm{N}_{scat}\times\eta,    
\end{equation}
where $\eta$ is the ratio of the energy enhancement of photons during scattering in 
the rest frame of the electron. 

\subsubsection{Unpolarized Photon}
According to \cite{Mcnamara2009}, the degree of linear polarization for initially unpolarized light
can be written as, 
\begin{equation}
P_{un} = \mathrm{sin}^2\theta\left[\eta + \eta^{-1} -1 + \mathrm{cos}^2\theta \right]^{-1}, 
\end{equation}
\begin{equation}
\label{eq:unP}
\mathrm{or}~\frac{\eta^2 + 1}{\eta}=\mathrm{sin}^2\theta \times (1+\frac{1}{P_{un}}),   
\end{equation}
where $\eta>1$ for inverse-Compton scattering. The Equation \ref{eq:unP} is simple and 
connects the thermal parameters of the electron cloud with the degree of polarization. The
same can be used for multiple scatterings, as Fig. 7 of \cite{Mcnamara2009} showed the variation of the degree of polarization with the average number of scatterings suffered by each photon.  
The left-hand side could further be simplified and written as 
\begin{equation}
\eta  \approx \mathrm{sin}^2\theta(1+\frac{1}{P_{un}}).
\end{equation}
Upon substitution in Equation \ref{eq:lag}, it will provide the relationship between {\it time lag and degree of polarization} for the {\it initially unpolarized photons} as 
\begin{equation}
\mathrm{t}_\mathrm{c}(\mathrm{E}_\mathrm{h}, \mathrm{E}_\mathrm{s}) = \frac{R}{c(1+\tau)}\frac{ln\left[\mathrm{N}_{scat}\times (1+\frac{1}{P_{un}})\times \mathrm{sin}^2\theta)\right]}{ln\left[1+4\Theta(1+4\Theta)\right]},
\label{eq:degup}
\end{equation}

where a spherical homogeneous electron cloud is considered. The initial unpolarized photons are particularly important for soft black-body emission from the standard accretion disk \citep{Chandra1950, SS73}.  

Alternatively, we can estimate the coronal radius using Eq. \ref{eq:degup}, which can be expressed 
in the following way
\begin{equation}
R = \mathrm{t}_\mathrm{c}(\mathrm{E}_\mathrm{h}, \mathrm{E}_\mathrm{s})\times c(1+\tau) \times \frac{ln\left[1+4\Theta(1+4\Theta)\right]}{ln\left[\mathrm{N}_{scat}\times (1+\frac{1}{P_{un}})\times \mathrm{sin}^2\theta)\right]},
\label{eq:radius}
\end{equation}
where, for a given set of polarization, time lag, and spectral parameters, one can estimate the 
radius of the radiatively active region. 

Again, the average number of scatterings suffered by the photons, or $\mathrm{N}_{scat}=\tau^2$, could be considered as $\tau>1$ and $\mathrm{N}_{scat}=\tau$ for $\tau<1$ \citep{PSS1983}. Thus, $\frac{\Delta \mathrm{N}_{scat}}{\mathrm{N}_{scat}} \sim \frac{2\Delta \tau}{\tau}$
could be assumed for the optically thick cases.

\subsubsection{Linearly Polarized}
Again, following \cite{Mcnamara2009}, we can write the degree of linear polarization as
\begin{equation}
P_{l} = 2\times \frac{1-\mathrm{sin}^2\theta \mathrm{cos}^2\phi}{\eta + \eta^{-1} - 2\mathrm{sin}^2\theta \mathrm{cos}^2\phi}, 
\end{equation}
where $\phi$ is the azimuthal angle. Simplification could be reduced to 
\begin{equation}
\frac{\eta^2 + 1}{\eta}  = \frac{2(1-\mathrm{sin}^2\theta \mathrm{cos}^2\phi)}{P_{l}} + 2\mathrm{sin}^2\theta \mathrm{cos}^2\phi.
\end{equation}
For inverse-Compton scattering ($\eta>1$) will yield
\begin{equation}
\eta \approx \frac{2}{P_{l}} + 2(1-\frac{1}{P_{l}})\mathrm{sin}^2\theta \mathrm{cos}^2\phi. 
\end{equation}
One can substitute in Equation \ref{eq:lag} to obtain the relationship between the {\it time lag} and {\it degree of 
polarization} for initially {\it linearly polarized} photons

\begin{equation}
\begin{split}
\mathrm{t}_\mathrm{c}(\mathrm{E}_\mathrm{h}, \mathrm{E}_\mathrm{s}) 
&= \frac{R}{c(1+\tau)} \times \\
&\quad \frac{\ln\left[2\mathrm{N}_{\mathrm{scat}} \left(\frac{1}{P_{l}} + \left(1 - \frac{1}{P_{l}}\right) \sin^2\theta \cos^2\phi \right)\right]}{\ln\left[1 + 4\Theta(1 + 4\Theta)\right]}
\end{split}
\label{eq:deglp}
\end{equation}

Linearly polarized incident photons could originate in the presence of non-thermal processes, such as synchrotron
radiation of a structured or unstructured magnetic field. However, the degree of linear polarization for unstructured magnetic fields will be much less compared to their structured counterparts \cite{DS1980, BKM2017}.  

\begin{figure}[t]
\vskip 0.5cm
  \centering
  \includegraphics[angle=0,width=0.5\textwidth,keepaspectratio=true]{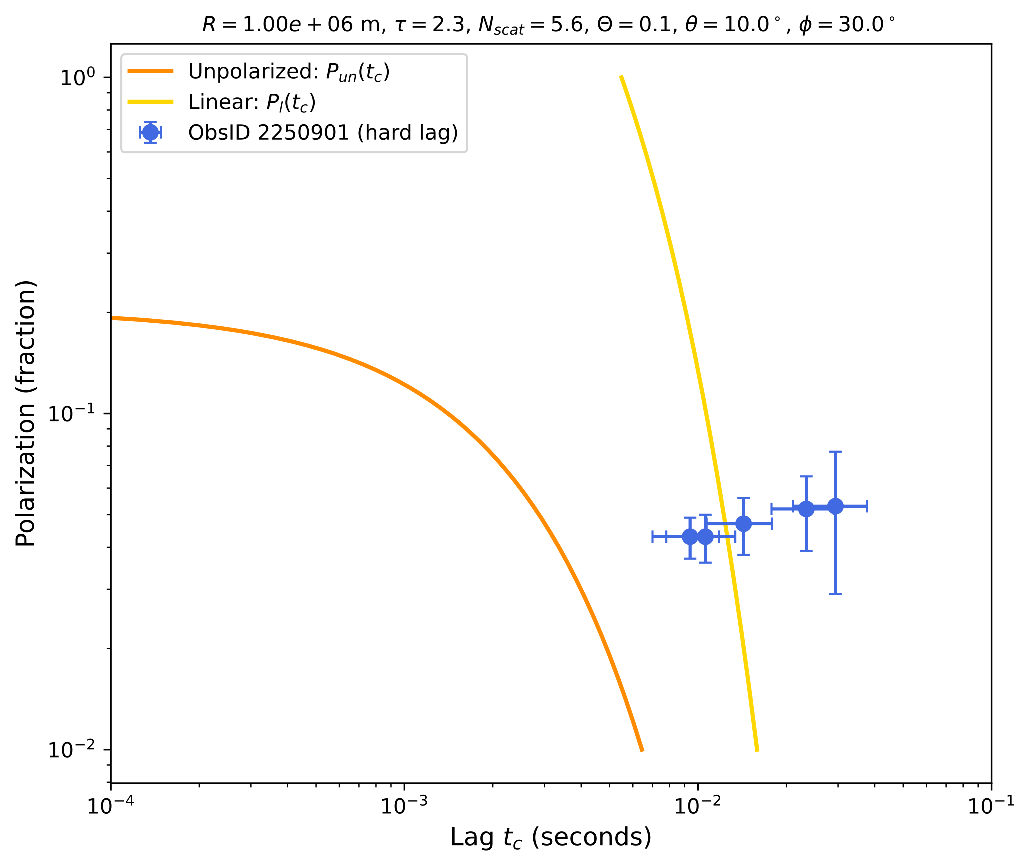}
	\caption{Polarization fraction is plotted as a function of time lag. The initially unpolarized photons are more likely to become depolarized with small time lags (fewer scatterings), whereas photons with structured polarization require many scatterings to become depolarized. Results from the IXPE observation on epoch 1 has been over-plotted for comparison. The region on the right side of the yellow line could indicate the presence of other physical processes contributing to the polarization of lagged hard X-rays.}
    \label{fig:pvl}
\end{figure}

Equations \ref{eq:degup} \& \ref{eq:deglp} suggest a depolarization effect due to the inverse Compton scattering. Fig. \ref{fig:pvl} shows the theoretical depolarization effect, while the observed polarization
increases with respect to energy along with the time lag and is presented in Figure \ref{fig:pds_papd_lag}. 

The relationships could be useful to detect the degree of X-ray polarization for previous and future non-polarimetric X-ray missions. 

\section{Conclusions}
\label{sec:conclusion}
\source~has provided an opportunity to study the spectral, temporal, and polarimetric properties of 
the black hole accretion simultaneously. We performed the timing and polarimetric observations using the {\it IXPE}
data, and spectral analysis was performed using the broadband X-ray data from {\it IXPE} and {\it Insight}/HXMT. Our study leads to the following major conclusions:
\begin{enumerate}
    
    \item We found a correlated energy-dependent time lag and degree of polarization in 
          the presence of a type-C QPO. The Spearman correlation coefficient was 0.8 for the \textit{IXPE} observation conducted on 7th September 2023. Moreover, we have observed that the correlation 
          depends on the Quality factor of the QPOs.

    \item The theoretical depolarization effect with inverse Compton scattering is explored; such an effect would yield anti-correlations between time lag and degree of polarization. Thus, the origin of high-energy polarization should be explored. Further explorations with {\it XPoSAT}/POLIX \citep{Paul2016} and {\it eXTP} \citep{extp2025} are essential to observe such a trend beyond $10$ keV. It should be noted that larger data sets, {\it IXPE} observations of QPOs, are essential to probe the lag-polarization relationships further.

\end{enumerate}

The probable causes for such variation of polarization may include the synchrotron self-Compton (SSC) \citep{Errando2024}, reflections close and distant to the black hole \citep{FRLW1989, PDM2024}, magnetic reconnection within the corona and jet region \citep{Zhang2020a}, and the shock-induced variation of polarization \citep{Allison1962, Fabas2011, Deng2017}.

\section*{Acknowledgements}
AC acknowledges the UPES Seed grant (UPES/R\&D/SoAE/25062025/23) for partially supporting this research. AC acknowledges the IUCAA visiting Associateship program. SSH acknowledges support from the Natural Sciences and Engineering Research Council of Canada (NSERC) through the Canada Research Chairs and Discovery Grants programs, and from the Canadian Space Agency. This work reports observations obtained with the Imaging X-ray Polarimetry Explorer ({\it IXPE}), a joint US (NASA) and Italian (ASI) mission, led by Marshall Space Flight Center (MSFC). The research uses data products provided by the {\it IXPE} Science Operations Center (MSFC), using algorithms developed by the {\it IXPE} Collaboration (MSFC, Istituto Nazionale di Astrofisica - INAF, Istituto Nazionale di Fisica Nucleare - INFN, ASI Space Science Data Center - SSDC), and distributed by the High-Energy Astrophysics Science Archive Research Center (HEASARC). 

\section*{Data Availability}
This work made use of public archival data from the {\it IXPE} mission, a project jointly maintained by MSFC, SSDC, INAF, and INFN. This has also made use of the data from the {\it Insight}-HXMT mission, a project funded by the China National Space Administration (CNSA) and the Chinese Academy of Sciences (CAS).


\bibliographystyle{mnras}
\bibliography{refs} 




\bsp	
\label{lastpage}
\end{document}